\documentclass[a4paper,aps,pre,twocolumn,groupedaddress,showkeys,showpacs,floats,floatats,floatfix]{revtex4-1}
\usepackage[utf8]{inputenc}
\usepackage[english]{babel}
\usepackage{graphicx}
\usepackage{dcolumn} 
\usepackage{bm}
\usepackage{natbib}
\usepackage{latexsym}
\usepackage{mathrsfs}
\usepackage{amssymb}
\usepackage{amsmath}
\usepackage{amscd}
\usepackage{color}
\usepackage{pifont}
\usepackage{pstricks,pst-node,pst-text,pst-3d}
\usepackage{verbatim}
\usepackage{ulem}
\usepackage[T1]{fontenc}
\usepackage{hyperref}
\hypersetup{
  colorlinks   = true, 
  urlcolor     = black, 
  linkcolor    = red, 
  citecolor    = blue 
}
\definecolor{aquamarine}{rgb}{0.5, 1.0, 0.83}
\definecolor{blue-violet}{rgb}{0.54, 0.17, 0.89}

\newrgbcolor{Red}{1.0 0.0 1.0}
\begin{document}
\title{Intermittent stickiness synchronization}
\author{Rafael M.~da Silva$^{1}$}
\email{rmarques@fisica.ufpr.br}

\author{Cesar Manchein$^{2}$}
\email{cesar.manchein@udesc.br}
\author{Marcus W.~Beims$^{1,3}$}
\email{mbeims@fisica.ufpr.br}
\affiliation{$^1$Departamento de F\'\i sica, Universidade Federal do Paran\'a, 
81531-980 Curitiba, PR, Brazil}
\affiliation{$^2$Departamento de F\'\i sica, Universidade do Estado de Santa 
Catarina, 89219-710 Joinville, SC, Brazil} 
\affiliation{$^3$Max-Planck Institut f\"ur Physik Komplexer Systeme, N\"othnitzer 
Stra\ss e 38, 01187 Dresden, Germany} 
\date{\today}
%
\begin{abstract}
This work uses the statistical properties of Finite-Time Lyapunov exponents (FTLEs) to 
investigate the {\it Intermittent Stickiness Synchronization} (ISS) observed in the mixed 
phase space of high-dimensional Hamiltonian systems. Full Stickiness Synchronization (SS) 
occurs when all FTLEs from a chaotic trajectory tend to zero for arbitrary long time-windows. 
This behavior is a consequence of the sticky motion close to regular structures which live in 
the high-dimensional phase space and affects all unstable directions proportionally by the 
same amount, generating a kind of {\it collective motion}. Partial SS occurs when at least one 
FTLE approaches to zero. Thus, distinct degrees of partial SS may occur, depending on the 
values of nonlinearity and coupling parameters, on the dimension of the phase space and on the 
number of positive FTLEs. Through filtering procedures used to precisely characterize the sticky 
motion, we are able to compute the algebraic decay exponents of the ISS and to obtain remarkable 
evidence about the existence of a {\it universal} behavior related to the decay of time 
correlations encoded in such exponents. In addition, we show that even though the probability to 
find full SS is small compared to partial SSs, the full SS may appear for very long times due to 
the slow algebraic decay of time correlations in mixed phase space. In this sense, observations 
of very late intermittence between chaotic motion and full SSs become rare events.   
\end{abstract}
%
\pacs{05.45.Ac,05.45.Pq}
\keywords{Stickiness, finite-time Lyapunov spectrum, Poincaré recurrences, 
synchronization.}
\maketitle

\section{Introduction}
\label{intro}
Synchronization in high-dimensional Hamiltonian systems has been defined as 
measure synchronization in Refs.~\cite{zanette99,vincent05}. In these works the 
authors use models consisting of two coupled maps. By starting two distinct 
initial conditions from the uncoupled system, which lead to a regular dynamics, 
they observe what happens to them by adding a small coupling between
the maps. A kind of synchronized (collective) motion appears named
{\it measure synchronization}. As the synchronization of dissipative
chaotic systems \cite{boca02,pecora15}, synchronization in generic
Hamiltonian systems is also an interesting issue since such systems
present a mixed phase-space dynamics which contains a rich variety of  
behaviors. However, it is important to mention that the
synchronization phenomenon observed in dissipative systems is not
possible in Hamiltonian system due to the Liouville's theorem that
prevents the full  collapse of the orbits to an invariant manifold,
since volume must be preserved in phase space.

For symplectic two-dimensional maps the chaotic component is clearly
separated from  the regular motion \cite{Lichtenberg,z08}. However in
higher dimensions the chaotic trajectory may visits ergodically the
whole phase space but, until this happens, it may suffer a dynamical
trapping (or sticky motion) \cite{Chir-Shep,Artuso} close to
quasi-regular structures. The effect of the sticky motion on the
chaotic trajectory can be classified in distinct {\it regimes}
\cite{Contopoulos,Malagoli}, defined by the spectrum of Finite-Time
Local Lyapunov Exponents (FTLLEs). When all FTLLEs are positive, the
regime is chaotic, when all are zero, we have an ordered regime. In
between we have semiordered regimes. Separating the dynamics in
distinct regimes, like a filtering procedure, not only a substantial
increase in the characterization of the sticky motion was achieved
\cite{RMS91}, but allowed to find a synchronized-like state, leading
to the {\it Intermittent Stickiness Synchronization} (ISS) discussed
in the present work. Essentially the ISS is characterized by the
intermittent behaviour between the chaotic motion and a kind of
transient measure synchronization generated by stickiness. Such
synchronized-like states due to stickiness were also detected some
years ago and classified as {\it common motion} \cite{cesar12}. A
somehow similar analysis allowed to synchronize drive and slave
coupled standard maps \cite{das16}. In this case, since the coupling
between the two maps is unidirectional, once the synchronized state is
reached, it does not change along the simulations. Such behavior
changes when the coupling interaction between the maps
is bidirectional, as considered in the present work, which uses global
(all-to-all) interactions.   

Since events with long times are associated to times for which the
trajectory was trapped to the nonhyperbolic components of the phase
space \cite{Artuso,Cristadoro,AltmannKantz,Shep2010,lrobk14,lbk16}, in
the present work the ISS decay is qualitatively described using the
time decay of the ordered regime in the case of coupled maps. To
mention, other alternatives approaches using FTLEs
\cite{GrassbergerKantz,fmv91,kk92,kk94,tl07,Viana,Harle} can be used
to characterize the phase space of conservative systems, with recent
applications using large deviation techniques 
\cite{Manchein,mam18,New} and the cumulants \cite{cesar12,cesar14}
of the FTLEs distribution.  

Recently, a methodology that uses time-series of local Lyapunov
exponents to define the above mentioned regimes of ordered,
semiordered, and totally chaotic motion was proposed, making it
possible to improve the characterization of stickiness in Hamiltonian
with few degrees of freedom \cite{RMS91} and non-Hamiltonian 
\cite{RMS92} systems. 

The present work uses such filtering procedure \cite{RMS91} to check
precisely the algebraic decay exponents of the ISS in
higher-dimensional Hamiltonian systems. This investigation is
motivated by the few amount of numerical studies related to weakly
chaotic properties and consequently the long time correlations
observed in  higher-dimensional mixed phase spaces of Hamiltonian
systems (at least for small and moderate number of homogeneously
coupled two-dimensional maps). We find that only the  full SS obeys a
power-law decay, while all other partial SSs decay
exponentially. Thus,  sticky effects from the semiordered regimes are
almost irrelevant for long time ISS decay. Additionally, the algebraic
decay exponent of full SS seems to be independent of
the (i) number of coupled maps (at least for a moderate number of
them), and (ii) the coupling intensities used here. Although it is
still under debate whether such an exponent persists in the
weak-coupling regime, our investigation corroborates with the claim
suggested in \cite{mam18} that predicts the existence of
{\it generic} decay exponent for time correlations $\chi \sim 0.20$
for Hamiltonian systems with few degrees of freedom which is smaller
than what says the conjecture proposed in \cite{Shep2010} to predict the
existence of an universal decay of Poincaré recurrences $\gamma
\approx 1.30-1.40$ (see also \cite{dbo90} for earlier work). The
corresponding relationship between these algebraic exponents is given
by the well-known equation $\chi=\gamma-1$. 

The plan of this paper is presented as follows. Section \ref{mod}
presents the coupled maps model used for the simulations. In
Sec.~\ref{method} the precise definition of ordered, semiordered, and
chaotic regimes is given, which leads to the definition of the
synchronized-like state, together with some numerical examples. While
in Sec.~\ref{ISS} the ISS decay is discussed qualitatively,
Sec.~\ref{conclusion} summarizes our conclusions. 

\section{The coupled-maps model}
\label{mod}

Consider the time-discrete composition $\mathbf{T} \circ \mathbf{M}$
of independent one-step iteration of $N$ symplectic $2$-dimensional
maps $\mathbf{M}=(M_1, \ldots, M_N)$ and a symplectic coupling $\mathbf{T}
=(T_1, \ldots, T_N)$. This constitutes a $2N$-dimensional Hamiltonian
system. For our numerical investigation we use the $2$-dimensional
Standard Map (SM):  
\renewcommand{\arraystretch}{1.3}
\begin{equation}
  \label{mp-acop1}
  \mathbf{M_i}\left(
  \begin{array}{c}
    p_i \\
    x_i \\
  \end{array}
  \right) = \left(
  \begin{array}{llll}
    p_i + K_i \sin(2\pi x_i) & \hspace{0.1cm} \mathrm{mod} \hspace{0.2cm} 1 \\
    x_i + p_i + K_i \sin(2\pi x_i) & \hspace{0.1cm} \mathrm{mod} \hspace{0.2cm} 1 \\
  \end{array}
  \right),
\end{equation}
\noindent and for the coupling 
\begin{equation}
  \label{mp-acop2}
  \mathbf{T_i} \left(
  \begin{array}{c}
    p_i \\
    x_i \\
  \end{array}
  \right) = \left(
  \begin{array}{llll}
    p_i + \sum_{j=1}^{N} \xi_{i,j} \hspace{0.05cm} \sin[2\pi (x_i - x_j)] \\
    x_i \\
  \end{array}
  \right),
\end{equation}
\noindent with $\xi_{i,j} = \xi_{j,i} = \frac{\xi}{\sqrt{N-1}}$ (all-to-all coupling). 
This system is a typical Hamiltonian benchmark tool with mixed phase space 
presenting all essential features expected in complex systems. It was studied 
in Refs.~\cite{AltmannKantz,AltmannThesis} using the Recurrence Time Statistic (RTS) 
and used in Ref.~\cite{RMS91} to propose the classification of Lyapunov regimes for 
improving stickiness characterization. In all numerical simulations we used 
nonlinearity parameters corresponding to a mixed phase space, namely 
$0.60\leq K \leq 0.65$ (see Fig.~\ref{psn1} which is discussed next). 

\begin{widetext}
$\quad$
\begin{figure}[!t]
  \centering
  \includegraphics*[width=0.99\columnwidth]{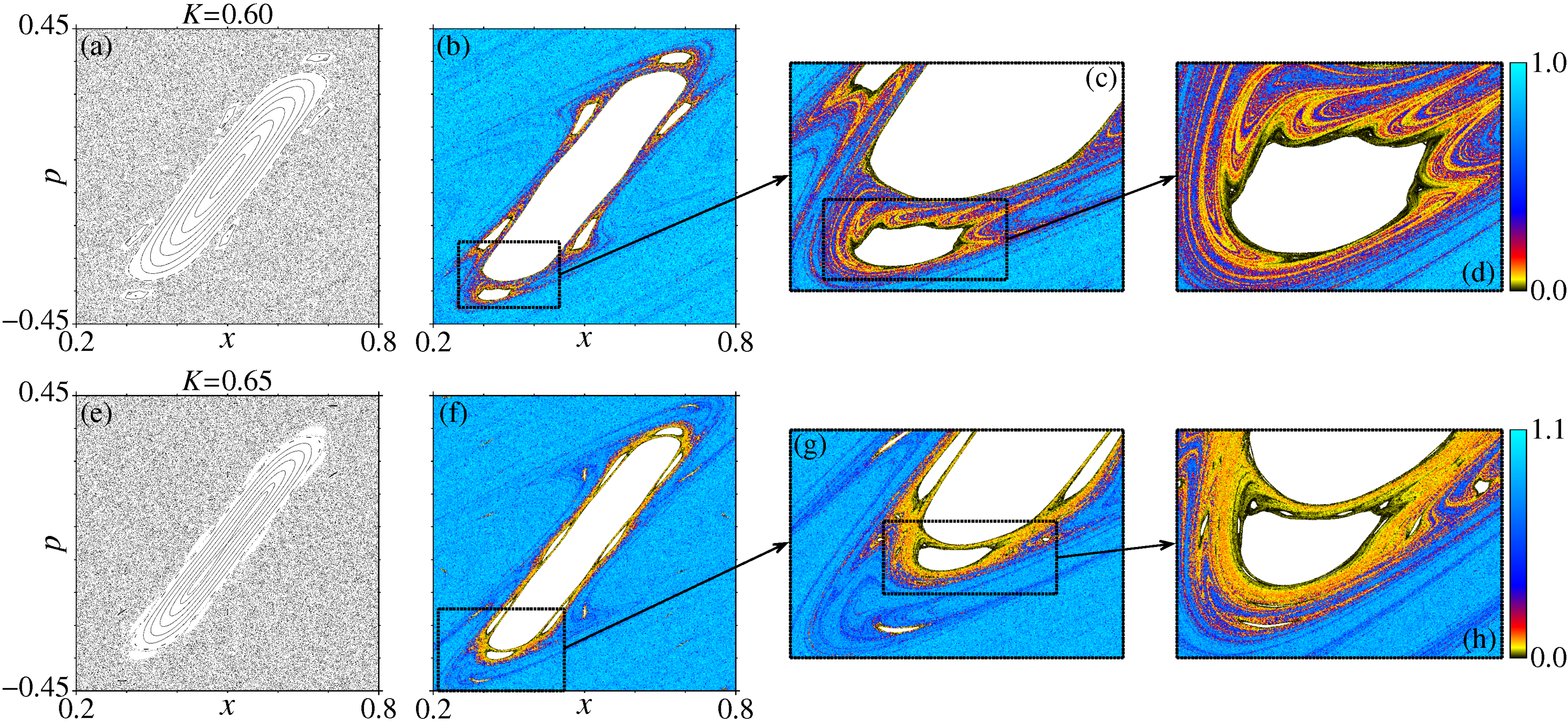}
  \caption{(Color online) Phase-space dynamics for the uncoupled case
    ($\xi=0$) using (a)-(d) $K=0.60$ and (e)-(h) $K=0.65$. In panels
    (a) and (e) we used $60$ initial conditions, each one iterated
    $8\times 10^5$ times. In (b) and (f) the colors (see the color
    bar) represent $\lambda_1^{(\omega)}$, for $\omega=100$, computed
    through a trajectory of $4\times 10^8$ iterations. Panels (c), (d)
    and (g), (h) are magnifications of the cases $K=0.60$ and $K=0.65$, 
    respectively.} 
  \label{psn1}
\end{figure}
\end{widetext}

\section{Definition of regimes and stickiness synchronization} 
\label{method}

The numerical technique uses the FTLLEs spectrum
$\{\lambda_{i=1\ldots  N}^{(\omega)}\}$ computed along a chaotic
trajectory during a window of size $\omega$, where
$\lambda^{(\omega)}_1 >
\lambda^{(\omega)}_2,\ldots,\lambda^{(\omega)}_N > 0$, and explores
temporal properties in the time series of $\{\lambda_i^{(\omega)}\}$
\cite{RMS91}. The window size $\omega$ has to be sufficiently small to
guarantee a good resolution of the temporal variation of the
$\lambda^{(\omega)}_i$'s, but sufficiently large in order to have a
reliable estimation (see
Refs.~\cite{GrassbergerKantz,Viana,Harle}). The sharp transitions
towards $\lambda_i^{(\omega)} \approx 0$ observed earlier motivates
the classification in regimes of motion
\cite{Contopoulos,Malagoli}. For a system with $N$ degrees of freedom,
the trajectory is in a regime of type $S_M$ if it has $M$ local
Lyapunov exponents $\lambda^{(\omega)}_i > \varepsilon_i$, where
{$\varepsilon_i \ll \lambda_i^{(\infty)}$} are small
thresholds. Thereby, $S_0$ and $S_N$ are {\it ordered} and {\it
  chaotic} regimes respectively. Regimes $S_i$ with $0<i<N$ are called
{\it semiordered}. For the computation of the FTLLEs we use the
traditional Benettin’s algorithm \cite{bggs80,wolf85}, which includes
the Gram–Schmidt re-orthonormalization procedure. On average, the
FTLLEs are in decreasing order. However, inversions of the order
({$\lambda_{i+1}^{(\omega)} >  \lambda_i^{(\omega)}$}) may happen for
some times $t$ and we have chosen to impose the order of
{$\lambda_i^{(\omega)}$} for all $t$.   

\subsection{The uncoupled case: $N=1$}

\begin{table}[!b]
\caption{\label{tab1} Values of $K_i$ used to couple the standard maps and the 
thresholds $\varepsilon_i$.} 
\label{t1}
\begin{center}
\begin{tabular}{|c|c|c|c|c|}
\hline
Value of $K_i$ & $N=2$ & $N=3$ & $N=4$ & $N=5$ \\ \hline
$K_1$ & 0.65 & 0.65 & 0.65 & 0.65 \\ \hline
$K_2$ & 0.60 & 0.63 & 0.64 & 0.64 \\ \hline
$K_3$ &  -   & 0.60 & 0.63 & 0.63 \\ \hline
$K_4$ &  -   &  -   & 0.62 & 0.62 \\ \hline
$K_5$ &  -   &  -   &  -   & 0.61 \\ \hline
Threshold &   &  &   &  \\ \hline
$\varepsilon_1$ & 0.10 & 0.10 & 0.10 & 0.10 \\ \hline
$\varepsilon_2$ & 0.05 & 0.08 & 0.08 & 0.08 \\ \hline
$\varepsilon_3$ &  -   & 0.05 & 0.06 & 0.07 \\ \hline
$\varepsilon_4$ &  -   &  -   & 0.04 & 0.06 \\ \hline
$\varepsilon_5$ &  -   &  -   &  -   & 0.04 \\ \hline
\end{tabular}
\end{center}
\end{table}

To get a better understanding of the involved complexity in the dynamics and the 
behavior of the FTLLEs, Fig.~\ref{psn1} displays the phase-space dynamics for a 
chaotic trajectory for one uncoupled ($\xi=0$) SM together with the corresponding 
positive FTLLE $\lambda^{(\omega)}_1$ (see color bar) for $\omega=100$. In this 
case the Lyapunov spectrum has only two Lyapunov exponents which, asymptotically, 
one is positive and the other one negative. Thus, only two regimes are observed: 
(i) the ordered one, if $\lambda_1^{(\omega)} < \varepsilon_1$ and, (ii) the 
chaotic one, if $\lambda_1^{(\omega)} > \varepsilon_1$. While the upper row in 
Fig.~\ref{psn1} presents the $K=0.60$ case, the lower row shows results for 
$K=0.65$. For both cases, the phase space has a large regular island located in 
the center, surrounded by higher order resonances. In Fig.~\ref{psn1}(a) we observe 
a $6$-order resonance and in Fig.~\ref{psn1}(e) a $8$-order resonance [better seen 
in Figs.~\ref{psn1}(b) and \ref{psn1}(f), respectively]. It is well known that 
additional higher-order resonances (not visible in the scale of these Figures) live 
around the island. These islands lead to the dynamical trapping which can be stronger, 
or not, depending on the topological structure of the islands. Such dependency becomes 
better visible when the positive FTLLE $\lambda^{(\omega)}_1$ is calculated for the 
trajectories. This is shown in colors in Figs.~\ref{psn1}(b) and \ref{psn1}(f) with 
some magnifications (see black boxes) shown respectively in Figs.~\ref{psn1}(c), 
\ref{psn1}(d) and \ref{psn1}(g), \ref{psn1}(h). We observe that, when approaching 
the island borders, the FTLLE decreases, as specified by the color 
bars in Figs.~\ref{psn1}(d), \ref{psn1}(h). A very complex behavior is evident and only 
motions very close to the regular islands have smaller FTLLEs. This suggests that these 
motions close to the regular islands will belong to the ordered motion.

\subsection{The coupled case: $N=2$}

\begin{figure}[!b]
  \centering
  \includegraphics*[width=1.0\columnwidth]{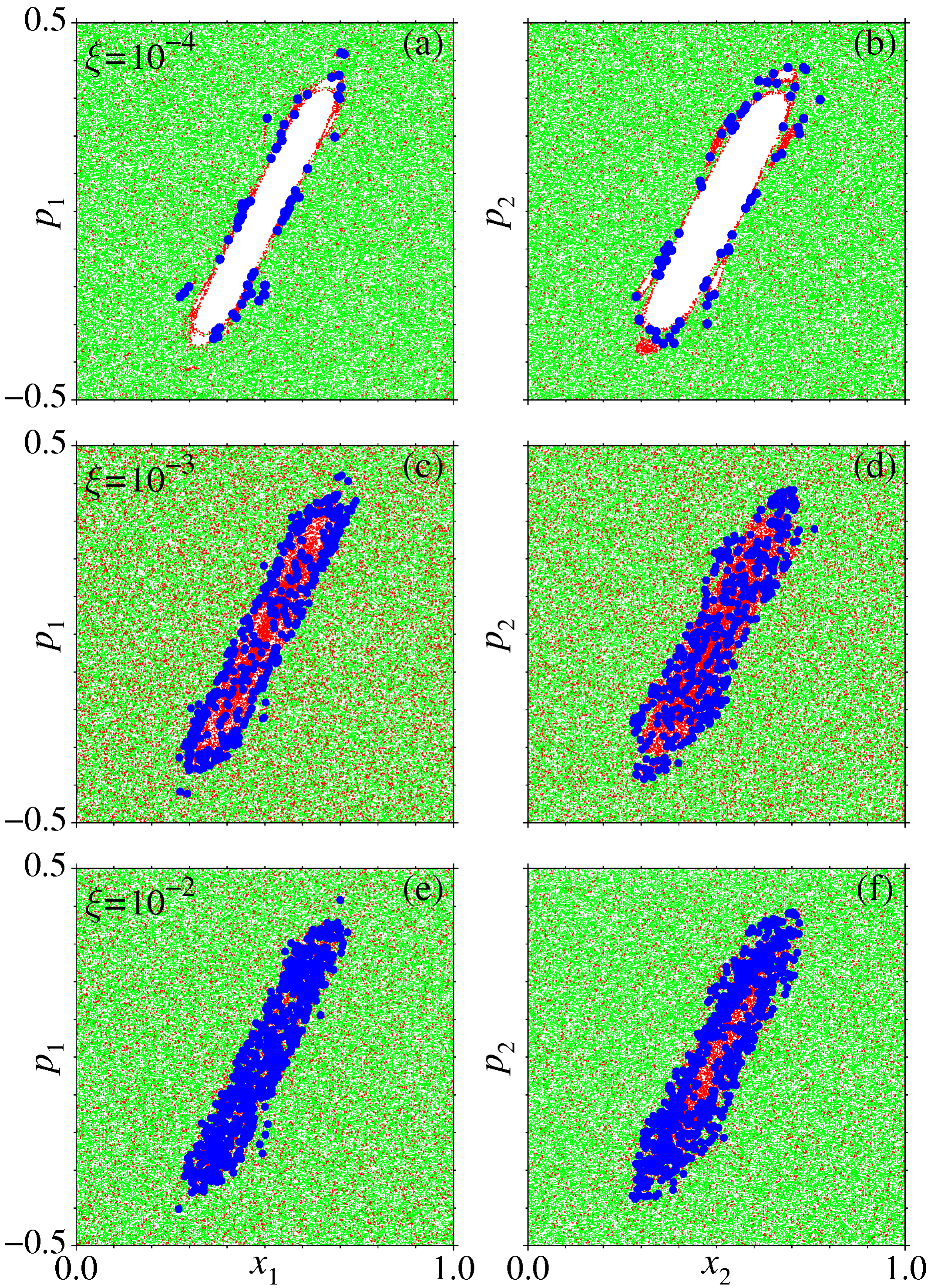}
  \caption{(Color online) Phase-space dynamics projected in $(x_1,p_1)$ [(a), 
  (c) and (e)] and in $(x_2,p_2)$ [(b), (d) and (f)] for different values of 
  $\xi$.}
  \label{psn2}
\end{figure}

A nice visualization of the regimes becomes clear when two coupled SM are analyzed 
in phase space, as shown in Fig.~\ref{psn2}. Different colors represent points 
$\boldsymbol{x}_{t}\in S_M$ belonging to regimes $S_0$ (blue circles), $S_1$ (red 
points), and $S_2$ (green points). These points were computed starting a single 
trajectory in the chaotic sea  of the coupled system and iterating it $10^7$ times. 
Table \ref{t1} presents the values of $K_i$ used in the simulations and the thresholds 
$\varepsilon_i$ used to define the regimes of motion. Blue circles are the points in 
phase space  ($x_1,p_1$) and ($x_2,p_2$) for which $\lambda^{(\omega)}_i < \varepsilon_i$, 
for $i=1,2$. The red color indicates points for which $\lambda^{(\omega)}_1 > 
\varepsilon_1$ and $\lambda^{(\omega)}_2 < \varepsilon_2$. Green points are used if both 
FTLLEs are larger than the respective thresholds $\varepsilon_i$. We observe in 
Fig.~\ref{psn2} that by increasing the value of the coupling strength $\xi$, the 
trajectory penetrates the regular domain from the uncoupled case where there is the 
island's hierarchy, inside which only $S_0$ and $S_1$ regimes occur. Thus, by increasing 
the coupling force between the maps, the number of points which induce sticky motion 
increase.

\subsection{The stickiness synchronization}

\begin{figure}[!t]
  \centering
  \includegraphics*[width=1.0\columnwidth]{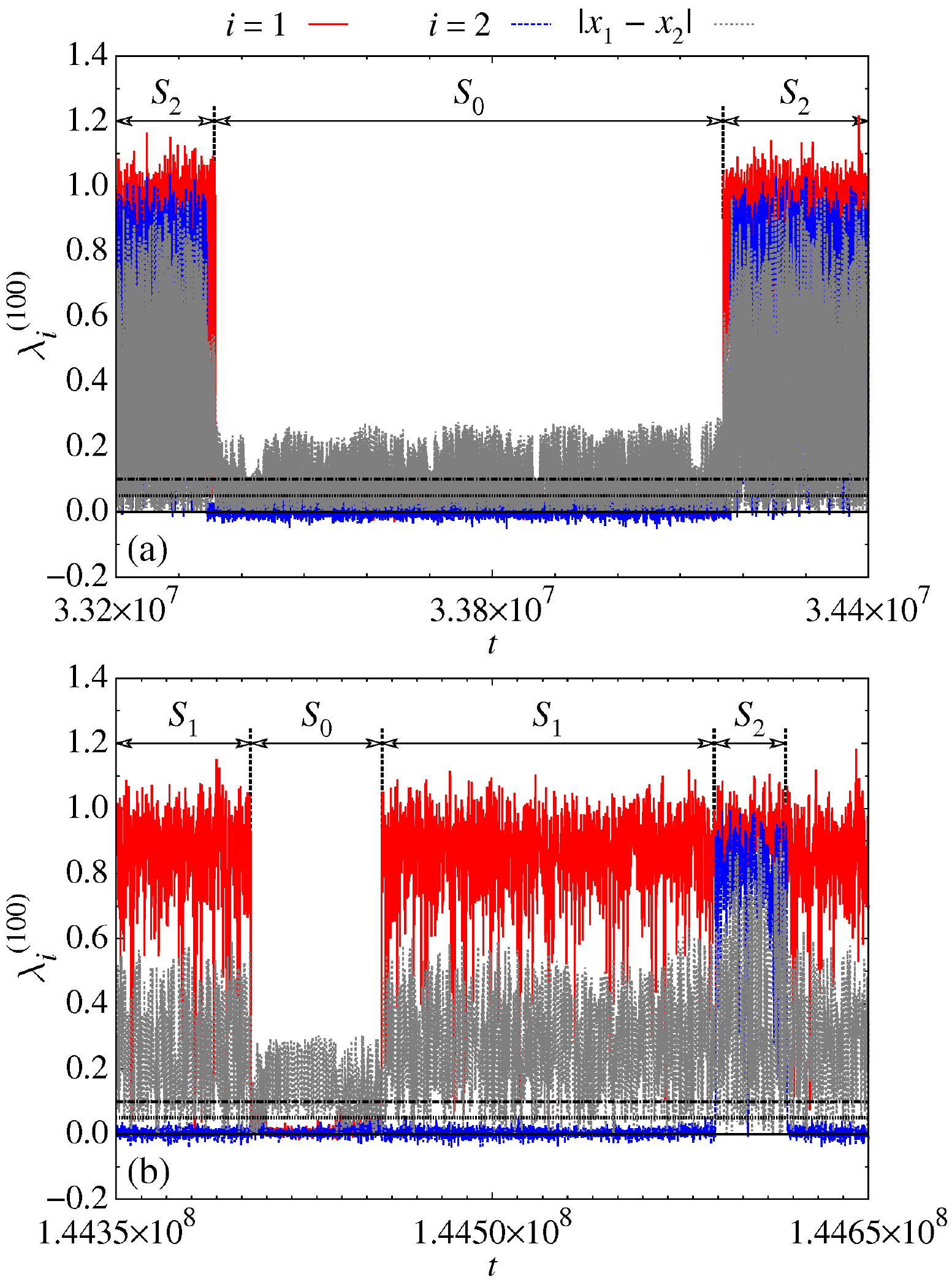}
  \caption{(Color online) Time series of the spectrum of FTLLEs $\{\lambda^{(
    \omega=100)}_i\}$, with $i=1,\ldots,N$ for the map (\ref{mp-acop1})–(\ref{mp-acop2}) 
    with $N=2$ and $\xi=10^{-3}$. In (a) and (b) the thresholds $\varepsilon_1=0.1$ and 
    $\varepsilon_2=0.05$ are represented by black dash-dotted and black dotted lines,
    respectively. The gray line indicates the distance $|x_1 - x_2|$ and shows the 
    synchronization of the maps $1$ and $2$ in the intervals of time for which the regime 
    $S_0$ occurs.}
  \label{sync}
\end{figure}

From above results it is easy to verify that for the ordered regime the position 
of coupled maps tend to be very close to the almost regular domains and to each 
other. This can be checked more precisely by determining, for $N=2$ for example, 
the distance $|x_1-x_2|$ as a function of time. This is shown by the gray color in 
Fig.~\ref{sync} for two distinct time windows. At a given time, the distance 
$|x_1-x_2|$ suddenly approaches zero, leading to an approximated synchronization of 
the positions of the coupled SMs. Since these positions are not exactly equal, we 
say to have a synchronized-like state. In both cases the synchronization occurs only 
for a finite-time window. Surprisingly these time windows match with those times for 
which the ordered regime $S_0$ is present. This can be checked in the same picture, 
where we plot simultaneously the two positive FTLLEs $\lambda^{(\omega)}_i$ as a 
function of time.  
Thus, synchronization of the position of the maps coincides with the full 
synchronization of the FTLLEs. For the regime $S_1$ we observe in Fig.~\ref{sync}(b) 
that the distance $|x_1-x_2|$ is more away from zero than this distance measured in 
the regime $S_0$, leading to a kind of ``weaker'' synchronization. In this case we 
say to have a {\it partial} synchronization, since only one FTLLE tends to zero. For
$S_2$ there is no synchronization.  

The relation between the position synchronization of the coupled maps shown above 
allows us to use the concept of {\it stickiness synchronization}. We have checked this 
relation for all $S_0$ regimes along a chaotic trajectory of length $t=10^9$.  
Since all regimes $S_M$ with $M<N$ are transient, and there is an intermittent transition 
between these regimes, we say to have the ISS. Our results for higher-dimensions can also 
be interpreted as the synchronization of FTLLEs. It is in fact a consequence of the 
synchronization of local expansion/contraction rates along all unstable/stable direction 
manifolds.

\section{Qualitative description of ISS}
\label{ISS}

Numerical techniques used to characterize the sticky motion can now be used to describe 
the qualitative behavior of the ISS decay in time. For this we use the consecutive time 
$\tau_M$ spent by a trajectory in the same regime $S_M$ \cite{RMS91}. During a 
trajectory of length $t_L$ we collected a series of $\tau_M$ and important results are 
obtained by analyzing the cumulative distribution of $\tau_M$, defined as: 
\begin{equation}
  \label{pcum-tm}
  P_{\mathrm{cum}}(\tau_M) \equiv \displaystyle \sum_{\tau'_M=\tau_M}^{\infty}
  P(\tau'_M).
\end{equation}

Applying this alternative procedure to obtain the $P_{\mathrm{cum}}(\tau_M)$, 
we are able to estimate the decay exponent for the recurrence times. This 
technique is much more appropriated to estimate such an exponent when compared 
to the former one, based on the cumulative distribution of Poincar\'e recurrence 
times [or (RTS)], since it remains unclear how to estimate the time scale over 
which a higher-dimensional system reaches its asymptotic regime under the process 
of weak Arnold diffusion \cite{Lichtenberg}. As this technique is based on a 
filtering process, we can select the events related to longer correlations and 
then obtain decay curves with several decades. In addition, there is no more 
dependence on the choice of recurrence set to obtain the RTS \cite{sam16}.  

\subsection{The uncoupled case: $N=1$}
\label{n1}

\begin{figure}[!t]
  \centering
  \includegraphics*[width=1.0\columnwidth]{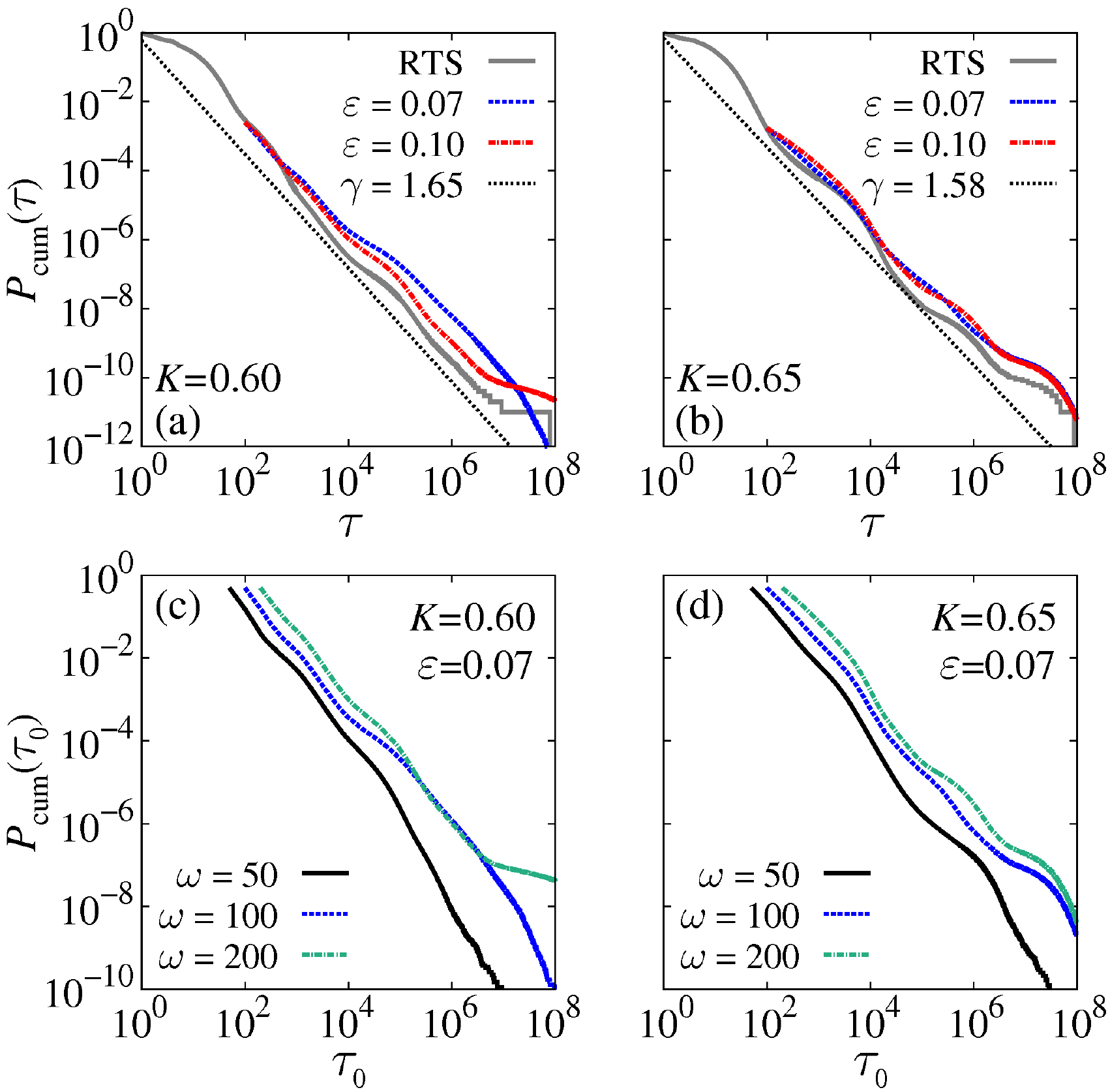}
  \caption{(Color online) Comparison between the filtering 
  method and the RTS (gray curves) for the uncoupled case ($\xi=0$) 
  {in (a) for $K=0.60$ and in (b) for $K=0.65$}. The blue and red curves 
  are the cumulative distribution $P_{\mathrm{cum}}(\tau_0)$ of consecutive times 
  $\tau_0$ in the regime $S_0$ {(normalized for convenience of scale)}, obtained 
  for $10^{11}$ values of $\tau_M$, using two different values of threshold 
  $\varepsilon$. The RTS $P_{\mathrm{cum}}(\tau)$ was obtained for $10^{12}$ 
  recurrences. {In (c) and (d) we compare $P_{\mathrm{cum}}(\tau_0)$ for 
  different values of $\omega$.}}
  \label{rts}
\end{figure}

To apply the filtering method we have to specify the threshold
$\varepsilon$ and the time window $\omega$. {Figures \ref{rts}(a) and 
\ref{rts}(b) compare} the cumulative distribution $P_{\mathrm{cum}}(\tau_M)$ for 
the regime $S_0$, obtained using $\omega=100$ and two values of $\varepsilon$,
with the RTS $P_{\mathrm{cum}}(\tau)$, both quantities calculated for
uncoupled SMs with two different values of $K$, specified in  
Fig.~\ref{rts}. For the determination of the RTS (plotted in gray) we {
have: (i) used a recurrence region in the chaotic component of
the phase space delimited by $0<x<1$ and $0.45<p<-0.45$, and (ii)
collected the lapses of time $\tau$ spent outside of the recurrence
region by a trajectory started inside of such predefined box.} Straight
lines in Fig.~\ref{rts} are consequences of the sticky motion.  We realize 
that while the usual RTS presents some oscillations as a function of $\tau$, 
leading to difficulties in the precise decay exponent, the filtering method 
{\it tends} 
to decrease such oscillations, mainly if the threshold $\varepsilon=0.07$
is used. These results show that, for practical implementations, the
thresholds can be defined as $\varepsilon_i 
\approx0.10\langle \lambda_i^{(\omega)}\rangle$, where $\langle \ldots
\rangle$ denotes the average over $t$, where $t=1,\ldots, t_L$. 

{It is important to define how sensitive these results are in relation 
to the time window $\omega$ used to calculate the FTLLEs. For this, we compare 
$P_{\mathrm{cum}}(\tau_0)$ obtained using $\omega=100$ and $\varepsilon=0.07$ 
[blue curves in Figs. \ref{rts}(a) and \ref{rts}(b)] with
$P_{\mathrm{cum}}(\tau_0)$  for $\omega=50$ and $\omega=200$, keeping the threshold 
$\varepsilon=0.07$. Figures \ref{rts}(c) and \ref{rts}(d) show this comparison for 
the cases $K=0.60$ and $K=0.65$, respectively, and the results demonstrate that 
even though the choice of $\omega$ may affect quantitatively the cumulative
distributions $P_{\text{cum}}(\tau_M)$, our conclusions about the algebraic decay 
obtained for the regime $S_0$ are not changed by oscillations around the chosen 
value $\omega=100$ } \cite{RMS91}. 

\subsection{The coupled cases: $N=2,3,4,5$}
\label{coupled}

\begin{figure}[!t]
  \centering
  \includegraphics*[width=1.0\columnwidth]{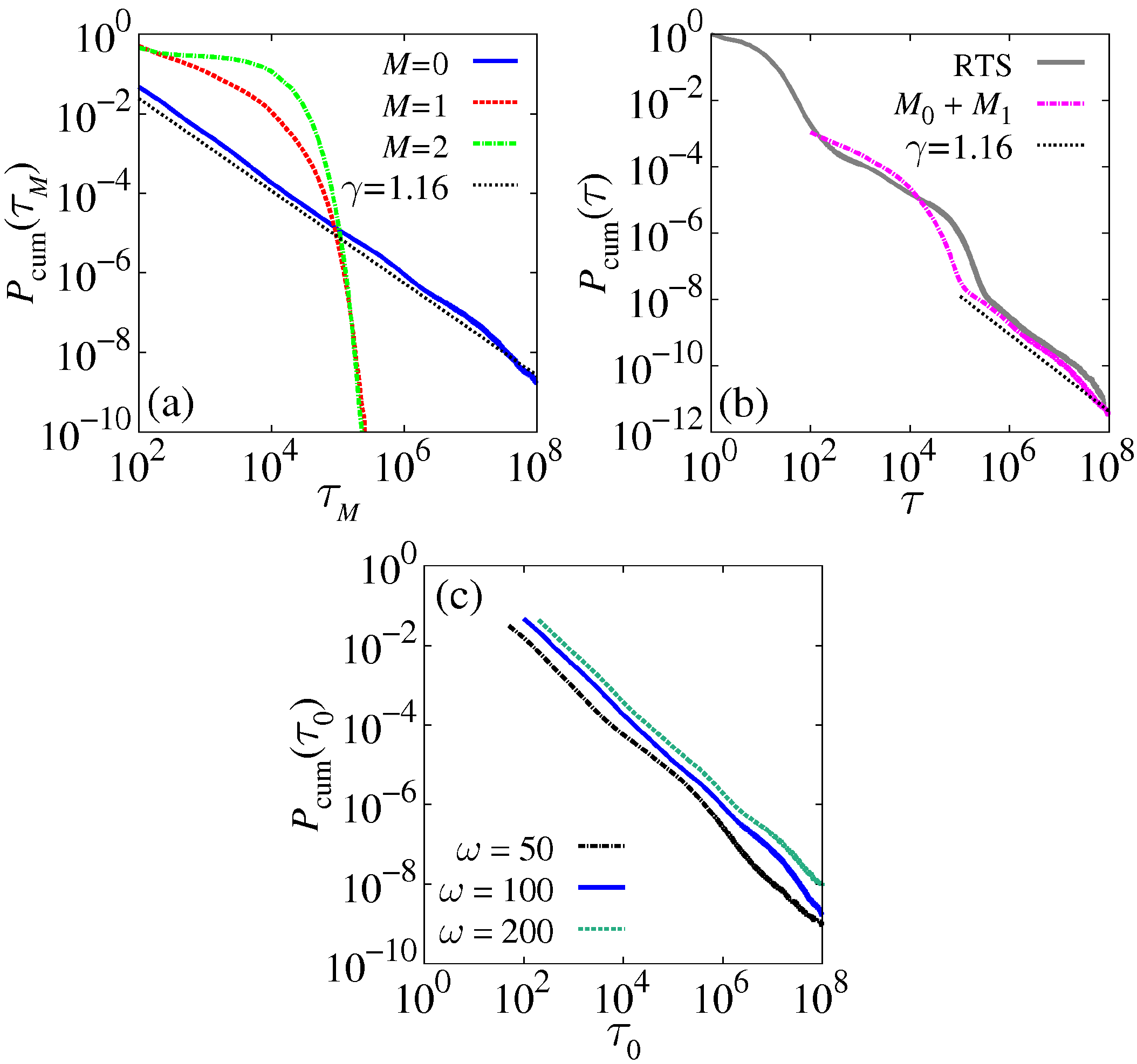}
  \caption{(Color online) (a) The cumulative distribution $P_{\mathrm{cum}}(\tau_M)$ 
  of times $\tau_M$ for the regime $S_M$ for the map (\ref{mp-acop1})–(\ref{mp-acop2}) 
  with $N=2$ and $\xi=10^{-3}$, obtained using $2\times 10^{10}$ values of $\tau_M$. 
  The values of $K_i$ and $\varepsilon_i$ used are indicated in the Table \ref{t1}. 
  {(b) Comparison between our method and the analysis based on RTS for the 
  case $N=2$ and $\xi=10^{-3}$. The result obtained combining the curves $M_0 + M_1$ 
  (normalized for convenience of scale) is equivalent to cumulative distribution 
  $P_{\mathrm{cum}}(\tau)$, obtained for $10^{12}$ recurrences. In (c) we compare 
  $P_{\mathrm{cum}}(\tau_0)$ for different values of $\omega$.}}
  \label{pcumN2}
\end{figure}

We start determining the  cumulative distribution $P_{\mathrm{cum}}(\tau_M)$ for 
the $N=2$ case for which we have regimes $S_0, S_1$ and $S_2$. This is shown in 
Fig.~\ref{pcumN2}(a) for coupling $\xi=10^{-3}$ and using values of $K_i$ 
and $\varepsilon_i$ according to the Table \ref{t1}. It shows that the only 
power-law decay occurs for the $S_0$ regime. Thus, while full SS occurs for $S_0$ 
regimes with a power-law decay of the $P_{\mathrm{cum}}(\tau_M)$, all other 
regimes have a chaotic component leading to an exponential decay. This indicates 
that only full FTLLEs synchronized states tend to occur for consecutive 
very long times, even though with small probability.

{
Looking at the distributions $P_{\mathrm{cum}}(\tau_M)$ in Fig. \ref{pcumN2}(a), 
we observe for the semiordered regime $M=1$ an exponential tail after an initial 
power-law decay with scaling $\beta \approx 0.5$ \cite{AltmannKantz}. When the full SS 
takes place ($M=0$), $P_{\mathrm{cum}}(\tau_0) \propto \tau_0^{-\gamma}$, with 
$\gamma=1.16$. As shown in Fig. \ref{pcumN2}(b), this scaling is compatible with the 
result obtained using RTS. However, the cumulative distribution 
$P_{\mathrm{cum}}(\tau_{0})$ provides a better characterization of algebraic decay 
(over several orders of magnitudes), which is essential when dealing with 
high-dimensional systems (which may contain different pre-asymptotic regimes) and 
for an accurate estimation of the stickiness exponent $\gamma$. In Fig. 
\ref{pcumN2}(c) we show that $P_{\mathrm{cum}}(\tau_{0})$ for the coupled case 
remains (qualitatively) the same for different values of $\omega$ around $\omega=100$.}

\begin{figure}[!t]
  \centering
  \includegraphics*[width=0.88\columnwidth]{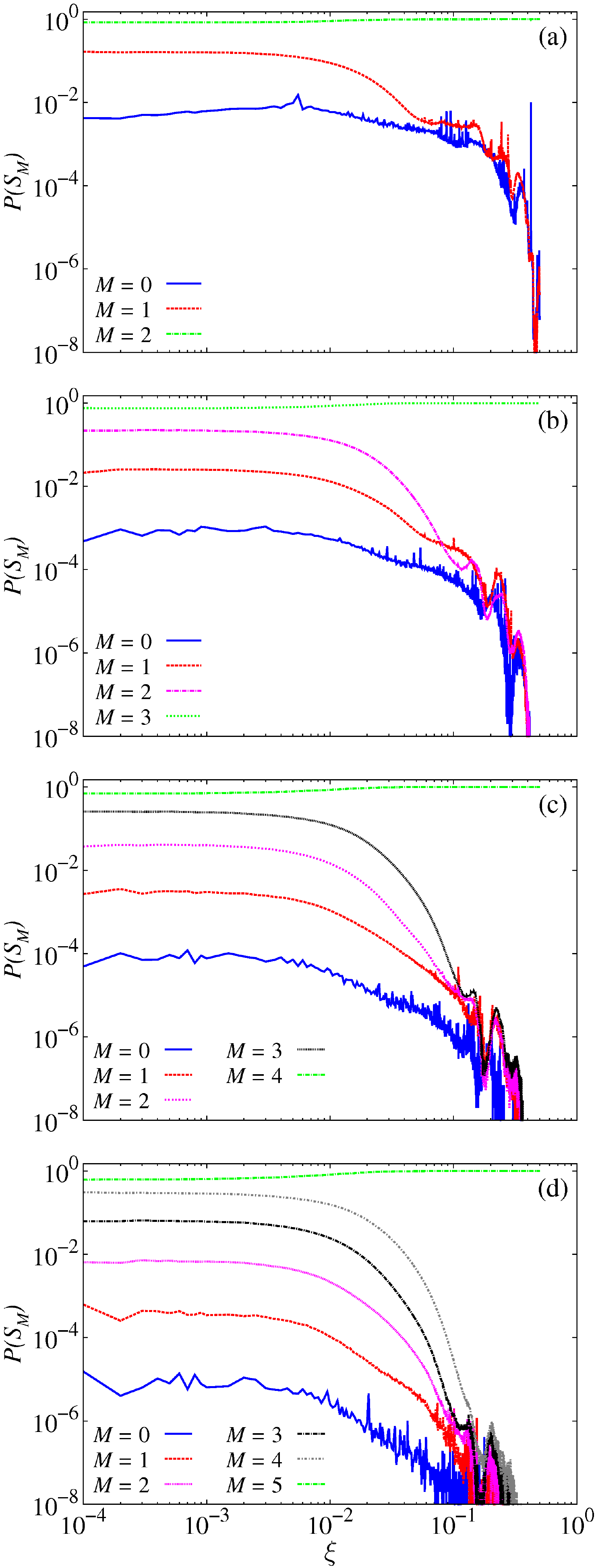}
  \caption{(Color online) Residence time in each regime $S_M$ using $\omega=100$
  and (a) $N=2$, (b) $N=3$, (c) $N=4$ and (d) $N=5$. For each value of $\xi$, 
  $P(S_M)$ was computed using a trajectory with length $t_L=10^{10}$. The values 
  of $K_i$ and $\varepsilon_i$ used in each case can be found in Table \ref{t1}.}
  \label{resid}
\end{figure}

Another very interesting quantity to be studied is the residence time $P(S_M)$ 
in each regime as a function of the coupling strength, defined by
\begin{equation}
\label{residence}
P(S_M) = \displaystyle \dfrac{1}{\beta} \displaystyle \sum_{t=0}^{t_L} 
\delta_{t \in S_M},
\end{equation}
\noindent where $\beta=t_L/\omega$ is the factor of normalization. In Eq. 
(\ref{residence}), $\delta_{t \in S_M} = 1$ if in time $t$ the trajectory is in
the regime $S_M$ and $\delta_{t \in S_M} = 0$ otherwise. The $P(S_M)$ is shown 
in Fig.~\ref{resid}(a) for $N=2$, in Fig.~\ref{resid}(b) for $N=3$, in Fig.
\ref{resid}(c) for $N=4$ and in Fig. \ref{resid}(d) for $N=5$. For
smaller couplings ($\xi\lesssim3\times10^{-2}$) the residence time decreases 
with  $M$, namely $P(S_N)>P(S_{N-1})>\ldots>P(S_M)>\ldots>P(S_1)>P(S_0)$. 
This means that  the probability to find the ordered regimes ($M=0$) is much 
smaller when compared to semiordered regimes ($M=1$) and so on. For larger couplings 
$\xi>10^{-1}$, the probability to find order and semiordered regimes has 
roughly the same values and tend to decrease until zero. Only the probabilities
of fully chaotic regimes $S_N$ remain for  larger values $\xi$.

From Figs.~\ref{pcumN2} and \ref{resid} we conclude that even though
the probability  to find  the full SS is small compared to the partial
SS, it can occur for very long times due to the power-decay found for
$P_{\mathrm{cum}}(\tau_0)$. In addition we mention that, in
distinction to usual synchronization {found in dissipative systems}, here 
the ISS tends to decrease for larger coupling between the maps
{\cite{prl74-95}}. 

{
\subsection{Characterizing the decay of synchronization: the ordered regime}
\label{s0}
}

\begin{figure}[!b]
  \centering
  \includegraphics*[width=1.0\columnwidth]{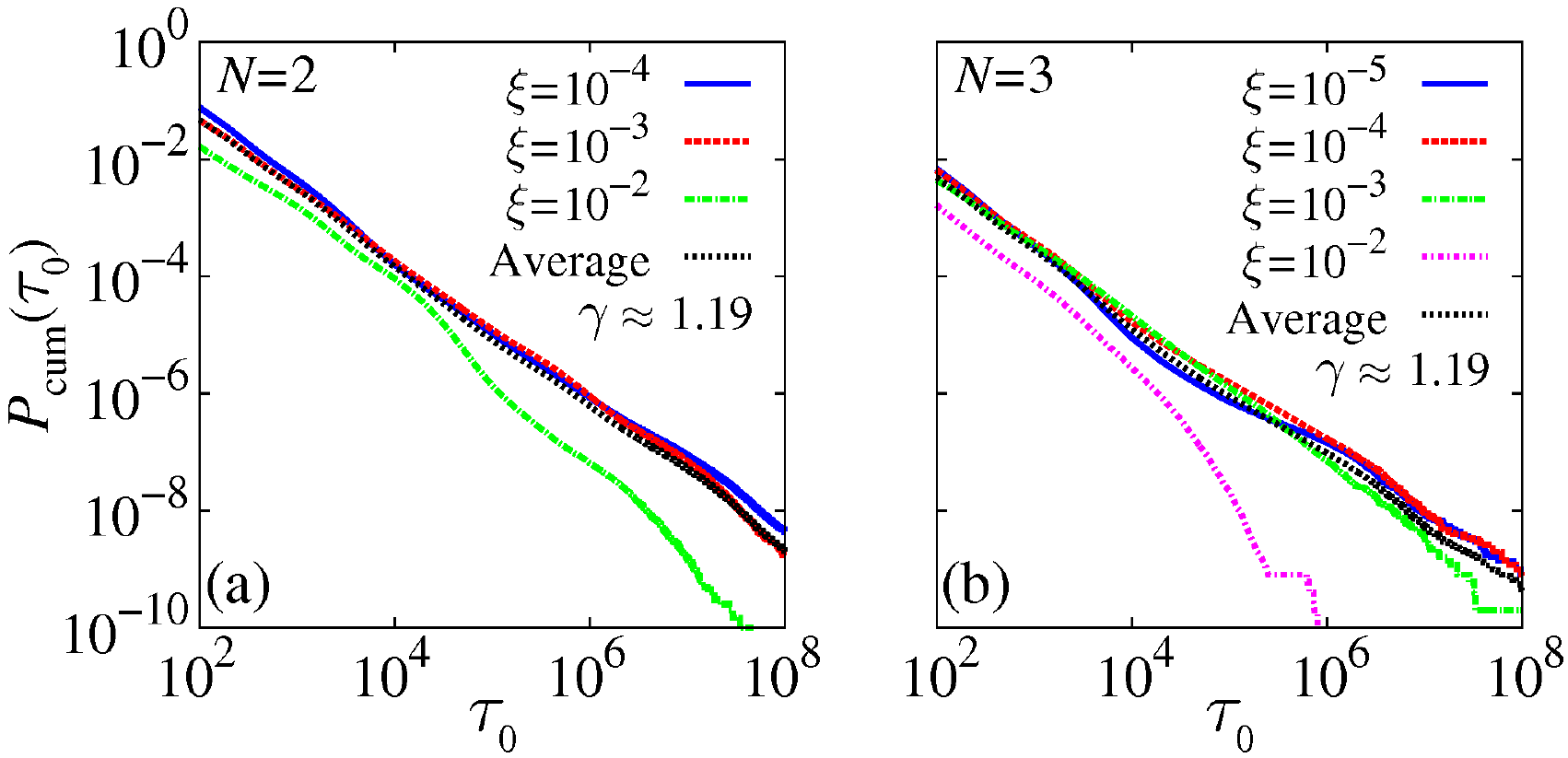}
  \caption{(Color online) The cumulative distribution {$P_{\mathrm{cum}}(\tau_0)$ 
  of consecutive times $\tau_0$ in} the regime $S_0$ using different values of
  coupling $\xi$ for (a) $N=2$ with $2\times 10^{10}$ values of
  {$\tau_M$} and (b) $N=3$ with $5\times 10^9$ values of
  {$\tau_M$}. The $K_i$ and $\varepsilon_i$ used in each case are
  indicated in the Table \ref{t1}.}
\label{taun23}
\end{figure}
{The next step is to precisely quantify the ISS decay for distinct
values of coupling $\xi$ between $N=2, 3, 4$ and $5$ SMs. For this we used only 
the regime $S_0$, which is directly related with the full synchronization between 
the positions $x_i$. As demonstrated in Figs. \ref{rts}(a), \ref{rts}(b) and 
\ref{pcumN2}(b), the decay of $P_{\mathrm{cum}}(\tau_0)$ provides an amazing 
characterization of the sticky motion and allows obtaining accurately the exponent 
$\gamma$, so that the RTS analysis becomes needless. The results of this study are} 
shown in Fig.~\ref{taun23}(a) for $N=2$ and in Fig.~\ref{taun23}(b) for $N=3$, 
{using} distinct values of $\xi$, as specified in the Figure. The 
black dotted line is the average over all couplings of each case and fitting this 
curve we obtain a power-law decay with well defined exponent $\gamma\approx 1.19$, 
observed for 6 decades. For $\xi=10^{-2}$, long-term trappings tend to disappear. 
It is worth to mention that the disappearance of the long-term sticky motion 
manifest itself in the increasing lack of data for $S_0$ as the coupling increases.  

\begin{figure}[!t]
  \centering
  \includegraphics*[width=1.0\columnwidth]{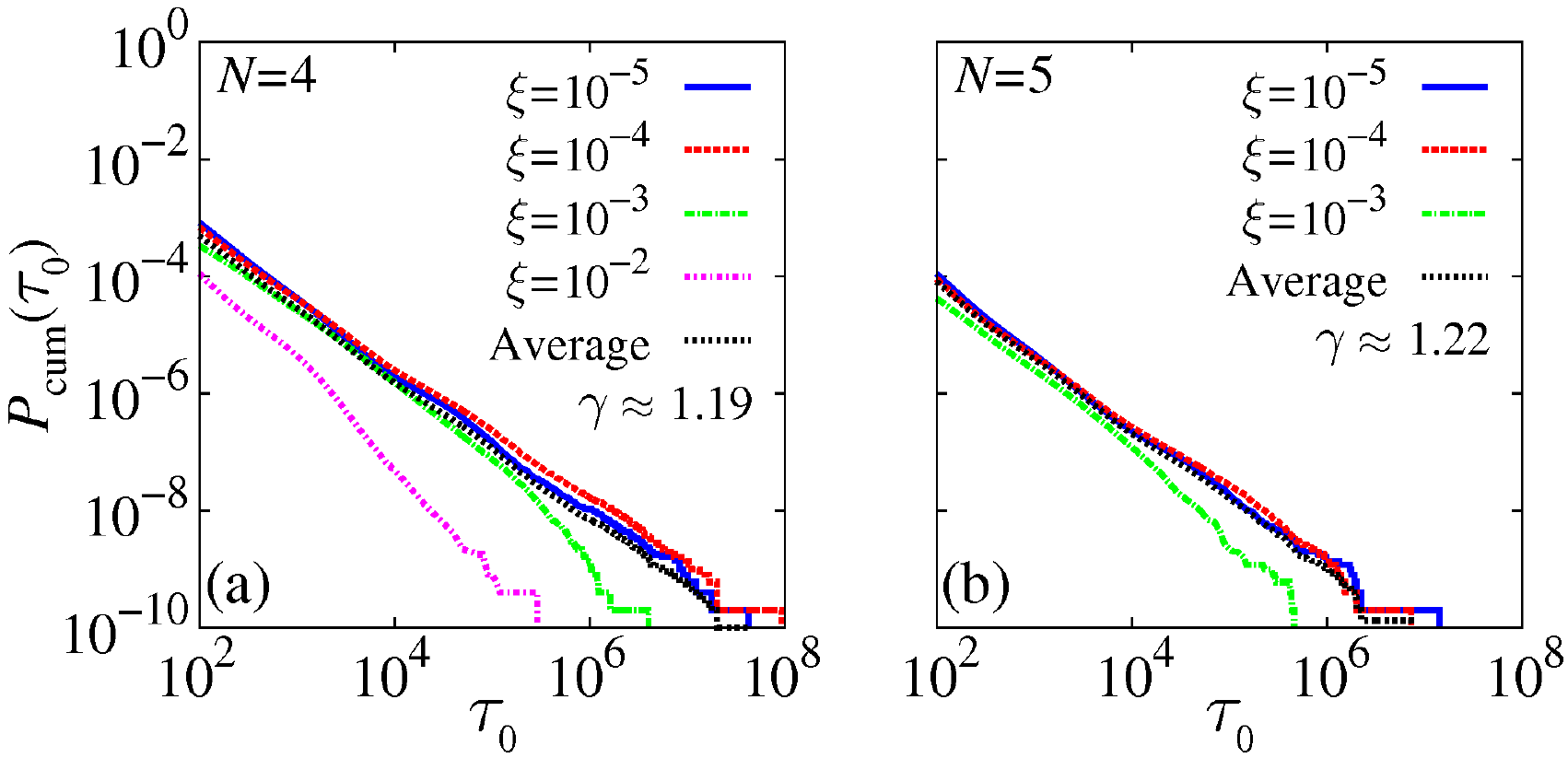}
  \caption{(Color online) The cumulative distribution {$P_{\mathrm{cum}}(\tau_0)$ 
  of consecutive times $\tau_0$ in} the regime $S_0$ using different values of
  coupling $\xi$ for (a) $N=4$ and (b) $N=5$, both cases collecting
  $5\times 10^9$ values of {$\tau_M$}. The $K_i$ and $\varepsilon_i$
  used in each case are indicated in  
  the Table \ref{t1}.}
  \label{taun45}
\end{figure}

To finish we would like to present results for $N=4$ and $N=5$.
Figure \ref{taun45} displays the cumulative distribution
{$P_{\mathrm{cum}}(\tau_0)$} for the regime $S_0$ and for distinct
coupling values, specified in the Figure. We observe that for values
of $\xi\le10^{-4}$ the exponent approaches $\gamma \approx 1.19$ for almost
$6$ decades in Fig.~\ref{taun45}(a), and $\gamma \approx 1.22$ in
Fig.~\ref{taun45}(b), values obtained fitting the black dotted line
that is the average over all couplings. The amount of long-term sticky
motion decreases for $\xi>10^{-4}$ in both cases. Again, this
manifests itself in the increasing lack of data for $S_0$. However,
since we still have at least three decades of power-law behavior, it
can still be characterized as sticky motion leading to the full SS.

\section{Conclusions}
\label{conclusion}

This work analysis qualitatively the Intermittent Stickiness Synchronization (ISS) 
decay in high-dimensional generic Hamiltonian systems. Such synchronization is 
generated by the regular structures on the chaotic trajectory, and can also be 
interpreted as the synchronization of FTLLEs. It is a synchronization of 
local expansion/contraction rates along all unstable/stable direction manifolds. We 
connect the intermittent motion between ordered, semiordered and  chaotic dynamical 
regimes with, respectively, the full, partial, and absence of synchronization 
generated by stickiness. By using the cumulative distribution of the
consecutive times $\tau_M$ spent in each regime $S_M$, we demonstrate
the ability of the recent proposed  filtering procedure \cite{RMS91}
to precisely characterize the ISS decay generated by the sticky
motion. We also show that even though the residence time in the full
SS state is small compared to the residence times in the partial SS
states, it may occur for {consecutive} very long times due to
the power-decay of the {$P_{\mathrm{cum}}(\tau_0)$}.   

In addition, our numerical results demonstrate that the algebraic
decay exponent tends to $\gamma \approx 1.20$ for higher-dimensional
systems. This is in completely agreement with the estimated decay
exponent of time correlations $\chi \approx 0.20$ (both exponents are
related by the well-know relationship $\chi=\gamma-1$) obtained in 
\cite{mam18} for $N=2,3$ symplectic maps interacting through a
nearest-neighbor coupling scheme. The estimated decay exponents in
these two works were obtained through three different approaches and
are somehow smaller than recent estimates \cite{Shep2010} (such 
observations suggest an universality conjecture, {at least for 
a moderate number of coupled Hamiltonian maps}).   

Future investigations can analyze a possible relation between the ISS
observed here and the hydrodynamic modes found in many body systems
\cite{Dellago96}. They present slow, long-wavelength behavior in the
tangent space dynamics. Besides, the properties of the covariant
Lyapunov vectors \cite{pol1,nor13,bg16,bg18} at the full SS might also
be promising from the theoretical point of view and applications. 
 
\vspace*{1cm}
\acknowledgments{ 
This study was financed in part by the Coordena\c{c}\~ao de
Aperfei\c{c}oamento de Pessoal de N\'\i vel Superior - Brasil (CAPES)
- Finance Code 001. C. M. and M. W. B.~thank CNPq (Brazil) for
financial support. The authors also acknowledge computational support
from Prof.~C.~M.~de Carvalho at LFTC-DFis-UFPR (Brazil).}


\end{document}